\documentstyle[aps]{revtex}

\newcommand{\bee}{\begin{equation}}
\newcommand{\ene}{\end{equation}}

\newcommand{\nats}{\mbox{${\rm I\!N }$}}
\newcommand{\intgs}{\mbox{${\rm Z\!\!Z }$}}
\def\beq{\begin{eqnarray}}
\def\eeq{\end{eqnarray}}
\newcommand{\nn}{\nonumber}
\newcommand{\cam}{{\cal M}}

\newcommand{\pa}{\partial}

\begin{document}
\title{Heat kernel asymptotics: more special case calculations} 

\author{Klaus Kirsten}
\address{Department of Physics and Astronomy, 
The University of Manchester, Oxford Road, Manchester M13 9PL, UK, 
email: klaus@a35.ph.man.ac.uk
\thanks{Invited contribution to the International 
Meeting on Quantum Gravity 
and Spectral Geometry, Naples, Italy, July 2--7, 2001.}
\thanks{ 
Research partially supported by EPSRC under Grant No GR/M45726 and by the 
MPI (Leipzig).}
\thanks{Address from 1. September 2001: Max-Planck-Institute 
for Mathematics in the Sciences,
Inselstr. 22-26, 04103 Leipzig, Germany.}} 

\maketitle
       
\begin{abstract}
Special case calculations are presented, which can be used to put 
restrictions on the general form of heat kernel coefficients for 
transmittal boundary conditions and for generalized bag boundary 
conditions.
\end{abstract}

\section{Introduction}
The heat equation asymptotics plays a prominent role in mathematics
\cite{gilk95b} and theoretical physics 
\cite{dewi75-19-295,dewi65b,eguc80-66-213,avra00b,espo94b,byts96-266-1,eliz94b}.
For manifolds without boundaries efficient schemes have been developed 
and the calculation of the volume part is nowadays nearly automatic
\cite{full88-310-583,avra91-355-712,ven98-15-2311}. When the manifold has
a boundary, suitable boundary conditions have to be imposed and the 
heat equation asymptotics receives additional contributions depending
on the boundary conditions considered
\cite{seel69-91-889,seel69-91-963}.
In recent years a conglomerate of methods has been proven to be very 
effective in the determination of this asymptotics for Laplace
type operators. The conglomerate of 
methods consists of functorial techniques, index theorems and special
case calculations. All ingredients are very agil in that they can be 
adopted to the specific boundary condition considered. A characteristic 
feature is that on its own none of the methods is able to determine the 
coefficients fully, but instead each determines only part of the complete 
answer. As a rule, the information obtained by each ingredient overlaps
with the information obtained from the others 
and this provides crucial cross checks for the calculation. Roughly 
one could say, that special case calculations are ``responsible'' 
for the group of extrinsic curvature terms, the index theorem allows to 
extract information on the gauge connection terms and functorial 
techniques, most prominently conformal transformation techniques
\cite{bran90-15-245}, determine 
whatever is left over. 
Employing this procedure, the heat equation asymptotics for the classic as 
well as for ``exotic'' boundary conditions has been determined. A summary 
is provided in the conference contribution \cite{gilk01naples}.
As further approaches let us mention 
\cite{avit91-8-603,avit91-8-1445,moss89-229-261,cogn90-241-381}.

The special case calculations that have been employed so far, consist of 
the following examples. (1) Dirichlet and Robin boundary conditions: 
scalar fields on the generalized cone $I\times {\cal N}$, $I=[0,1]$, ${\cal N}$
a Riemannian manifold \cite{bord96-37-895,bord96-182-371,kirs98-15-5,dowk01-42-434}. (2) Mixed boundary conditions: spinor fields and p-forms on the 
generalized cone \cite{dowk99-7-641,bran99-563-603}. (3) Oblique 
boundary conditions: again the generalized cone, in addition  
$B^2 \times T^{D-2}$ with the two-ball $B^2$ and the $(D-2)$-dimensional 
torus $T^{D-2}$ \cite{dowk99-16-1917}. (4) Spectral boundary conditions:
generalized cone plus $B^2 \times {\cal N}$ \cite{dowk99-7-641,dowk99-242-107,gilk00u}.

In the present contribution we will present two further special case 
calculations, one on a different manifold and one on the ball with 
different boundary conditions. 
First we provide a calculation
on a manifold obtained when glueing together a hemisphere and a ball.
This example determines crucial information for transmittal boundary conditions
\cite{gilk01-601-125}. Second we consider the Dirac operator on the ball with
boundary conditions as they occur in gauge theories in Euclidean bags. These 
boundary conditions involve an angle $\theta$, which is a substitute for 
introducing small quark masses to drive the breaking of chiral symmetry
\cite{hras84-245-118,wipf95-443-201}. An analysis of the associated 
heat equation asymptotics has barely started \cite{wipf95-443-201,gilk01u}.

\section{Hemisphere ball glued together}
The first example we present is of relevance in the context of transmittal
boundary conditions. To define transmittal boundary conditions, see 
e.g.~\cite{gilk01-601-125}, consider the $d$-dimensional manifold 
$\cam = \cam^+ \cup_\Sigma \cam^-$, which is the union of two 
compact manifolds $\cam^\pm$ along their common boundary $\Sigma$. 
Let $D^\pm$ be Laplace type operators on $\cam^\pm$ written invariantly 
as 
\beq
D^\pm = -g^{ij} \nabla_i^\pm \nabla_j^\pm - E^\pm .\nn
\eeq
The operator $D= (D^+ , D^-)$ acts on a pair $\phi = (\phi^+ , \phi^- )$.
The transmittal boundary condition is defined by the operator
\beq
{\cal B}_U \phi=\{\phi^+|_\Sigma-\phi^-|_\Sigma\}\oplus 
  \{(\nabla^+_{m^+}\phi^+)|_\Sigma
   + (\nabla^-_{m^-} \phi^-)|_\Sigma   +U\phi^+|_\Sigma\},
\label{naples1a}
\eeq
where 
$\nabla^\pm_{m^\pm}$ is the respective exterior normal derivative 
on $\cam^\pm$ to the boundary $\Sigma$.

Let us consider a specific example of the above situation and we start with the
two-dimensional case. 

Let ${\cal M}^+= H^2$ be 
the unit hemisphere and ${\cal M}^-=B^2$ the two-ball.
On the hemisphere we consider the ``massive'' Laplacian
\beq
D^+ = -\Delta_{H^2} + \frac 1 4 . \nn
\eeq
Adding the curvature term $(1/8) R = (1/4)$ has the advantage that the 
eigenvalues $\lambda^2$ of $D^+$ are complete squares. In detail, the 
eigenfunctions are 
\beq
\phi_{hemis} (\theta , \varphi ) = N_h 
      e^{im\varphi } P_{\lambda-1/2} ^{-|m|} (\cos \theta ) , 
       \quad       m\in \intgs , \nn
\eeq
with the associated Legendre functions $P_\nu ^\mu (x)$ and $N_h$ a 
normalisation constant. The values of $\lambda$ have to be fixed 
by boundary conditions.

On the ball we simply take
\beq
D^- = - \Delta_{B^2} , \nn
\eeq
with the well known eigenfunctions
\beq
\phi_{disc} (r,\varphi ) = N_d e^{im\varphi} J_{|m|} (\lambda r) ,\quad
m\in\intgs . \nn
\eeq
Here, $J_\nu (x)$ are the Bessel functions and $N_d$ is a normalisation 
constant.

Next we glue the hemisphere and the disc together along their boundary,
which is a circle. As a specific example of the transmittal boundary 
conditions (\ref{naples1a}) 
we impose that the eigenfunctions of $D^+$ and $D^-$ 
as well as their normal derivatives agree along the circle,
this is we choose $U=0$. 

Matching the eigenfunctions eliminates one of the normalization 
constants,
\beq
N_d J_{|m|} (\lambda ) = N_h P_{\lambda -1/2} ^{-|m|} (0) , \nn
\eeq
so 
\beq
N_d = N P_{\lambda-1/2} ^{-|m|} (0) , \quad
N_h = N J_{|m|} (\lambda ) . \nn
\eeq
Next we match the normal derivatives. The natural normal 
derivatives in the example are the exterior normal derivatives 
$(\partial / \partial r)$ and $(\partial / \partial \theta )$. In these 
the condition reads 
\beq
0 = \frac \pa {\pa \theta } \phi _{hemis} 
+\frac \pa {\pa r} \phi _{disc}
,\nn
\eeq
and this gives the implicit eigenvalue equation
\beq
0&=& \lambda J_{|m|} ' (\lambda ) P_{\lambda-1/2} ^{-|m|} (0) 
    - J_{|m|} (\lambda ) \frac d {dx}
            P_{\lambda -1/2} ^{-|m|} (x) \left|_{x=0}.
 \right. \label{naples1}
\eeq
In the following we will simplify the notation to 
\beq
\frac d {dx} P_{\lambda-1/2} ^{-|m|} (x) \left|_{x=0} = 
      \right. \frac d {dx} P_{\lambda-1/2} ^{-|m|} (0).\nn
\eeq
Once this implicit eigenvalue equation (\ref{naples1})
is known, the procedure developed 
in \cite{bord96-37-895} can be used in order to determine the heat equation
asymptotics for the specific problem at hand.
Starting point of the procedure is the contour integral representation
\beq
\zeta (s) = \sum_{m=-\infty} ^\infty \int\limits_\gamma \frac
      {dk}{2\pi i} (k^2 + M^2) ^{-s} 
      \frac \pa {\pa k}   
   \ln \left[ k J_{|m|} ' (k) P_{k-1/2} 
        ^{-|m|} (0) - J_{|m|} (k) \frac d {dx} P_{k-1/2} ^{-|m|} (0) 
       \right] , \label{naples2} 
\eeq
where the contour $\gamma$ encloses counterclockwise all real positive 
zeroes of equation
(\ref{naples1}) and 
where a small mass $M^2$ (as a infrared regulator put to zero later) has 
been introduced. 

The next step is to deform the contour to the imaginary axis, obtaining
\beq
\zeta (s) = \frac{\sin (\pi s)} \pi  \sum_{m=-\infty} ^\infty
       \int\limits _M ^\infty dk \,\, (k^2 - M^2) ^{-s} \frac \pa 
     {\pa k}      
        \ln \left[ k I_{|m|} ' (k) P_{ik-1/2}
        ^{-|m|} (0) - I_{|m|} (k) \frac d {dx} P_{ik-1/2} ^{-|m|} (0)
       \right], \label{naples1aa}
\eeq
where 
$P_\nu^\mu (x)= P_{-\nu -1}^\mu (x)$, this implies
$P_{ik -1/2}^{-|m|}  (x) = P_{-ik -1/2}^{-|m|} (x)$, has been used 
\cite{grad65b}. 

As explained in great detail in \cite{bord96-37-895}, 
the heat equation asymptotics is encoded in the
asymptotic behavior of the implicit eigenvalue equation (\ref{naples1}).
For the Bessel functions everything needed is provided by Olvers 
uniform asymptotic expansions. For $m\to \infty$ and $z=k/m$ fixed, 
they read \cite{olve54-247-328,abra70b},
\beq
I_m (mz) &\sim & \frac 1 {\sqrt{2\pi m}} 
      \frac{e^{m\eta}}{(1+z^2)^{1/4}}  
   \left[ 1+\sum_{k=1}^\infty \frac{u_k (t)} {m^k}\right],\label{olve1}\\
I_m ' (mz) &\sim & \frac 1 {\sqrt{2\pi m}} 
 \frac{e^{m \eta} (1+z^2)^{1/4}} z
         \left[ 1+\sum_{k=1}^\infty \frac
         {v_k (t)} {m^k} \right] , \label{olve2}
\eeq
with $t=1/\sqrt{1+z^2}$ and $\eta = \sqrt{1+z^2} + \ln [z / (1+\sqrt{1+z^2})]$.
The polynomials are defined  recursively through
\beq
u_{k+1} (t) &=& \frac 1 2 t^2 (1-t^2) u_k ' (t)
  + \frac 1 8 
   \int\limits_0^t d\tau \,\, (1-5 \tau^2) u_k (\tau ) ,\nn\\
v_k (t) &=& u_k (t)
         + t (t^2-1) \left[ \frac 1 2 u_{k-1} (t) + 
    t u_{k-1}' (t) \right] ,\nn
\eeq
starting with $u_0 (t) =1$. The first few coefficients are listed in 
\cite{abra70b}, higher coefficients are immediate to obtain by the above
recursions using a simple computer program. 

The relevant information
for the Legendre functions is \cite{grad65b}
\beq
P_{\lambda-1/2} ^{-m} (0) &=& 
           \frac{2^{-m} \sqrt \pi} {\Gamma (3/4 + 
        (m+\lambda )/2) \Gamma (3/4 + (m-\lambda ) /2 ) } , \label{lege1}\\
\frac d {dx} P_{\lambda-1/2} ^{-m} (0) &=&  
     -\frac{2^{-m+1} \sqrt \pi } {\Gamma (1/4 + (m+\lambda) /2 ) 
      \Gamma (1/4 + (m-\lambda) /2 ) } . \label{lege2}
\eeq
Further expansion of the $\Gamma$-functions is achieved using 
\cite{grad65b} 
\beq
\ln \Gamma (z)  \sim  z\ln z - z -\frac 1 2 \ln z + \ln \sqrt{2\pi} 
       + \sum_{k=1} ^\infty \frac{B_{2k}} {2k (2k-1) z^{2k-1}} , \nn
\eeq
with the Bernoulli numbers $B_{2k}$. We use this in the slightly different
form
\beq
& &\hspace{-.5cm}\Gamma (u+a) \sim 
\exp\left(u\ln u -u + 
     (a-1/2) \ln u 
  + \ln \sqrt {2\pi} 
       +\sum_{l=1} ^\infty \frac{ \tilde \gamma_l (a) } {u^l} 
        \right)  ,\nn
\eeq
respectively 
\beq
 \frac 1 {\Gamma (u+a)} \sim \exp\left(
-u\ln u +u - (a-1/2) \ln u 
      -\ln \sqrt{2\pi} \right)
\left( 1 + \sum_{l=1} ^\infty \frac{\gamma_l (a) } {u^l}\right) , \nn
\eeq
with the $\gamma_l (a)$ easily determined by an algebraic computer program.

These expansions are used for $m\neq 0$ after substituting $k\to km$ in the 
integral (\ref{naples1aa}); 
a suitable choice turns out to be $u= m (1+ik) /2$. As known from 
the ball, $m=0$ needs special treatment and a simple large argument 
consideration is sufficient.

Without going into the details of this calculation, as to be expected the sum
over $m$ produces Riemann zeta functions and a typical expression for the
asymptotic contribution is
\beq
A_l (s) = 2 \frac{\sin (\pi s)} \pi \zeta _R (2s+l) \int\limits_0 ^\infty 
 dk \,\, k^{-2s} \frac \pa {\pa k} D_l (k) , \nn
\eeq
where, e.g.,
\beq
D_1 (k) = \frac{-2 -3k^2 + \sqrt{1+k^2} } {24 (1+k^2)^{3/2}} , \quad 
D_2 (k) = -\frac {k^4} {32 (1+k^2) ^3} .\nn
\eeq
The $k$-integrals are standard and residues and values of 
$\zeta (s)$ are easily determined. As a pleasant feature let us mention that
the whole calculation can be and has been fully automised with a simple
algebraic computer program.

The transition from two dimensions to arbitrary dimension $d$ is surprisingly
easy once the suitable organising functions are known. On ball and 
sphere settings the best organisation is obtained via the use of 
the Barnes zeta function 
\cite{barn03-19-426,barn03-19-374,dowk94-35-4989,bord96-182-371},
\beq
\zeta_{{\cal B}} (s,b) = \sum_{\vec m =0} ^\infty 
 \frac 1 {(b+m_1 + ... + m_d )^s } 
    = \sum_{l=0}^\infty {l+d-1 \choose d-1 } (l+b)^{-s} . \nn
\eeq
In the present context this is seen as follows. On the hemisphere, 
eigenfunctions can be represented 
as Gegenbauer polynomials \cite{erde55b}. The 
relevant dependence on the normal coordinate $\theta$ is 
\beq
P_{\lambda -1/2} ^{-l-(d-2)/2} (\cos \theta ) , \quad
l\in \nats_0,\nn
\eeq
where again a coupling 
\beq
E^+ = \frac {d-1} {4d} R^+ = \frac 1 4 (d-1)^2 
\eeq
has been included. The radial dependence
on the ball is
\beq
\frac {J_{l+(d-2)/2} (\lambda r) } {r^{(d-2)/2}} , \nn
\eeq
such that 
\beq
\frac \pa {\pa r}  \frac {J_{l+(d-2)/2} (\lambda r) } {r^{(d-2)/2}}
    = \lambda J_{l+(d-2)/2} (\lambda r) -\frac {d-2} 2 
         J_{l+(d-2)/2} (\lambda r) .\nn
\eeq
The angular tangential dependences of both solutions agree. 
Eq.~(\ref{naples1}) suggests, that the most convenient choice to proceed
is the implicit eigenvalue equation
\beq
0= \lambda J_{l+(d-2)/2} ' (\lambda ) P_{\lambda -1/2} ^{-l-(d-2)/2} 
     (0) 
 - J_{l+(d-2)/2}(\lambda ) \frac d {dx} 
     P_{\lambda -1/2} ^{-l-(d-2)/2}  (0) .\label{naples3}
\eeq
The index $m$ is replaced by $l+(d-2)/2$
and the degeneracy 
\beq
d(l) = (2l+d-2) \frac{(l+d-3)!}{l! (d-2)!} \nn
\eeq
of the spherical harmonics has to be taken into account.
Thus
final answers can be found
from the two dimensional result once the Riemann zeta function is 
replaced by a sum of 
the Barnes zeta function \cite{dowk94-35-4989,bord96-182-371},
\beq
    \sum_{l=0}^\infty d(l) \left( l+\frac{d-2} 2 \right)^{-2s} 
      =\zeta _{{\cal B}} (2s, (d-2)/2) + \zeta _{{\cal B}} (2s, d/2 ) .\nn
\eeq
Note, however, that eq.~(\ref{naples3})
does {\it not} correspond to matching the normal derivatives at the 
boundary, but instead they are assumed to have a jump described
by $U= (d-2)/2$,
\beq
\left( \frac \pa {\pa r} \phi _{disc} \right) \left|_{r=1}  
   + \left( \frac \pa {\pa \theta } \phi _{hemis} \right) \left|_{\theta 
      = \pi/2} 
       = -\frac {d-2} 2 \phi_{disc} 
          \right|_{r=1} \right.     . \nn
\eeq
In summary, the meromorphic structure of the zeta function for the 
$d$-dimensional problem is completely clear and can be used to calculate
the leading heat kernel coefficients. Subtracting the volume contributions
of ${\cal M}^+$ and ${\cal M}^-$, the ``boundary part'' of some leading
coefficients is
\beq
a_1 &=&0 ,\nn\\       
a_2 &=& \frac{(4-d) 2^{-d} } {3\Gamma (d/2)} , \nn\\
a_3 &=& \frac{ (d-5) (d-3) \sqrt \pi 2^{-d} } {64 \Gamma (d/2) } ,\nn\\
a_4 &=& \frac{2^{-d}} { 180 \Gamma (d/2) }  
 \left( - \frac{250} 7 + \frac{2839} {42} d -\frac{191} 7 d^2 
      + \frac{ 61} {21} d^3 \right) .\nn
\eeq
These results can and have been used to put restrictions on the general form 
of the coefficients for transmittal boundary conditions
\cite{gilk01-601-125}. To exemplify the procedure consider the leading 
coefficients. Let $K^\pm$ be the extrinsic curvature on $\Sigma$ as 
induced from $\cam^\pm$. Invariance theory shows that for $f$ a localizing 
test function
\beq
a_1 ^\Sigma &=& \int\limits_\Sigma dy \,\, c_1 f 
 \mbox{Tr }({\bf 1}) , \nn\\
a_2 ^\Sigma &=& (4\pi)^{-d/2} \frac 1 6 \int\limits_\Sigma dy \,\,
   \mbox{Tr}\left( e_1 f (K^+ + K^-) {\bf 1} 
   + e_2 (f^+ _{;m^+} + f^- _{;m^-}) {\bf 1} 
    + e_3 f U \right). \nn
\eeq
For the hemisphere ball example, we have $K^+ =0$, $K^- = d-1$, 
$U=(d-2)/2$, and derive the equations
\beq
c_1 &=& 0 , \nn\\
(d-1) e_1 + \frac{d-2} 2 e_3 &=& 4-d , \nn
\eeq
and so $e_1 =2$, $e_3 = -6$. This information, as well as the remaining 
one $e_2 = 0$, is easily obtained by different
means, but for the higher coefficients the special case input is crucial. 

\section{Generalized Euclidean bag boundary conditions}
In theories of Euclidean bags, chiral symmetry breaking is triggered
by imposing the boundary condition 
\beq
0 &=& \Pi_- \psi \left|_{\pa\cam}\right. \nn\\
  &:=& \frac 1 2 \left( 1+ie^{\theta \Gamma^5 }\Gamma^5 \Gamma^m \right) 
                \left|_{\pa\cam}\right. \label{naples2a}
\eeq
on the spinor field $\psi$. For $\theta =0$ this is a well studied problem
which can be understood as a mixed boundary problem of an associated 
Laplace type operator $D$. Let $e_j$ be a $d$-bein system,
$\Gamma_{jkl}$ the Christoffel symbols relative to the orthonormal frame and
$\Gamma^j$ the $\Gamma$-matrices projected along this frame. Let $P$ be 
the Dirac operator on $\cam$,
\beq
P = -i \Gamma^j \nabla_j,\nn
\eeq
where $\nabla_j = e_j + \omega_j$ is the covariant derivative with the spin
connection
\beq
\omega_j = -\frac 1 4 \Gamma_{jkl} \Gamma^k \Gamma^l . \nn
\eeq
Consider the associated second order problem for $D=P^2$ with domain
\beq
\mbox{domain}(D) = \{ \psi \in C^\infty (V) : \Pi_- \psi 
  \left|_{\pa\cam}
 \oplus  \Pi_- \left( - i \gamma^j \nabla_j \right)
                 \psi \right|_{\pa\cam} =0 \} .\nn
\eeq
For $\theta =0$, this leads to the mixed problem
\beq
\Pi_- \psi \left|_{\pa\cam}\right. \oplus 
     (\nabla_m - S ) \Pi _+ \psi \left|_{\pa\cam}\right. =0 ,
\label{naples2b}
\eeq
where 
\beq
\Pi_+ = \frac 1 2 \left( 1-i\Gamma^5 \Gamma^m \right) \nn
\eeq
and 
\beq
S = -\frac 1 2 K \Pi_+ .\nn
\eeq
In order to derive eq.~(\ref{naples2b}), it is crucial that 
$[\Pi _- , \Gamma^a ] =0$, see \cite{bran92-108-47}, a relation that does not
hold for $\theta \neq 0$. For the case $\theta =0$ 
the special case calculation on the ball has been performed in 
\cite{dowk96-13-2911,dowk99-7-641} and subsequently used to restrict the 
general form of the coefficients for mixed boundary conditions 
\cite{bran99-563-603}. It is the aim of this section, to generalize the 
special case calculation to $\theta \neq 0$ and to use the results in order
to determine the $a_1$-coefficient for a general manifold. The structure
of the higher coefficients is considerably more difficult due to the 
non-commuting $\Gamma$-matrices involved in the invariance theory.
However, work along the lines shown here is in progress \cite{gilk01u}.

We start the analysis by solving the eigenvalue problem on the
$d$-dimensional ball. The eigenspinors of the Dirac operator on the ball 
have the form \cite{dowk96-13-2911}
\beq
\psi_\pm ^{(+)} = \frac C {r^{(d-2)/2}} 
    { i J_{n+d/2} (kr) Z_+ ^{(n)} (\Omega )\choose 
              \pm  J_{n+(d-2)/2} (kr) Z_+ ^{(n)} (\Omega )} , 
   \label{naples2c}\\
\psi_\pm ^{(-)} = \frac C {r^{(d-2)/2}}
    {\pm  J_{n+(d-2)/2} (kr) Z_- ^{(n)} (\Omega )\choose
              i  J_{n+d/2} (kr) Z_- ^{(n)} (\Omega )} ,
   \label{naples2d}
\eeq
with $n\in \nats_0$. Here, $Z_\pm ^{(n)}(\Omega )$ are the spinor modes on 
the sphere, see \cite{camp96-20-1}, and $C$ is a normalisation constant.

The condition (\ref{naples2a}), 
\beq
\Pi_- \psi \left|_{\pa \cam} \right. = \frac 1 2 
\left(
\begin{array}{cc}
    1 & -i e^\theta \\
  i e^{-\theta} & 1 
\end{array}
\right) ,\nn
\eeq
is easily applied to (\ref{naples2c}) and (\ref{naples2d}), and the implicit
eigenvalue equations read
\beq
J_{n+d/2} (k) \mp e^\theta J_{n+d/2-1} (k) =0 \nn
\eeq
for $\psi_\pm ^{(+)}$, and 
\beq
J_{n+d/2} (k) \pm e^{-\theta } J_{n+d/2-1} (k) =0 \nn
\eeq
for $\psi_\pm ^{(-)}$. Suitably combined, we write these as 
\beq
J_{n+d/2-1}^2 (k) - e^{-2\theta} J_{n+d/2}^2 (k) &=& 0 , \label{naples2e}\\
J_{n+d/2-1}^2 (k) - e^{2\theta} J_{n+d/2}^2 (k) &=& 0 . \label{naples2f}
\eeq
Clearly it is sufficient to deal with eq.~(\ref{naples2e}), the 
contributions  from (\ref{naples2f}) follow by 
replacing $\theta \to -\theta$.

For convenience we introduce $\nu = n+(d-2)/2$. To analyse the zeta function
we start again with the contour integral representation
\beq
\zeta (s) = \sum d(\nu ) \int\limits _\gamma \frac{dk}{2\pi i} 
   k^{-2s} 
   \frac \pa {\pa k} \ln \left( J_\nu ^2 (k) - 
  e^{-2\theta } J_{\nu +1} ^2 (k) \right) , \nn
\eeq
where $d(\nu )$ is the degeneracy associated with the implicit eigenvalue 
equation (\ref{naples2e}) and $\gamma$ again encloses counterclockwise
the positive real zeroes of (\ref{naples2e}). 
(For convenience, we omit writing the infrared cutoff $M^2$. The 
procedure in principle is as in the previous section.) Denoting by $d_s$ the
dimension of spinor space, $d_s = 2^{d/2}$, and taking into account 
the sphere eigenspinor degeneracies, one finds
\beq
d (\nu ) = \frac 1 2 d_s {d+n-2 \choose n} .\nn
\eeq
Proceeding in the manner described, we shift the contour to the imaginary
axis to find
\beq
\zeta (s) = \frac{\sin \pi s} \pi \sum d(\nu ) \int\limits _0 ^\infty 
 dk \,\, k^{-2s} 
  \frac \pa {\pa k} \ln \left( I_\nu^2 (k) + e^{-2\theta } I_{\nu +1} 
   ^2 (k) \right) .\label{naples3a}
\eeq
To simplify the analysis of the asymptotic behavior of the integrand it 
is convenient to rewrite (\ref{naples3a}) in terms of Bessel functions 
involving only one index. For this purpose we use \cite{grad65b}
\beq
I_{\nu +1} (z) = I_\nu ' (z) -\frac \nu z I_\nu (z) , \nn
\eeq
which allows to write
\beq
\zeta (s) 
&=& \frac {\sin \pi s} \pi \sum d (\nu ) \nu^{-2s} \int\limits_0^\infty
dz \,\, z^{-2s} \times\nn\\
& &\frac \pa {\pa z} \ln \left(  
 e^{-\theta } {I_\nu ' }^2 (z\nu ) + \left[ e^{\theta} + \frac 
        {e^{-\theta } } {z^2} \right] I_\nu^2  (z \nu )
 -\frac 2 z e^{-\theta } I_\nu (z\nu ) I_\nu ' (z\nu ) \right),
\nn
\eeq
when irrelevant factors in the logarithm are neglected. 

The uniform asymptotics is completely determined by eqs.~(\ref{olve1}) and
(\ref{olve2}). Here we will concentrate only on the terms contributing
to the heat equation coefficients $a_0$ and $a_1$. It is easy to see, 
that the relevant pieces from the argument of the logarithm are
\beq
(...)= \frac{e^{2\nu \eta}}{2\pi \nu} 
 \frac{e^{-\theta} (1+z^2)^{1/2}}{z^2} 2 (1-t) 
 \left\{1+\frac {1+t} 2 \left[ e^{2\theta} -1 \right] \right\} 
     +   \mbox{irr.}. \label{naples4a}
\eeq
The contribution to $\zeta (s)$ resulting from the first line
has already been dealt with in the calculation for 
the case $\theta =0$ and it reads \cite{dowk99-7-641}
\beq
\zeta_1 (s) =  \frac{d_s}{4\sqrt{\pi}} \frac{\Gamma (s-1/2)} {\Gamma (s+1)} 
     \zeta_{{\cal B}} (2s -1, d/2-1) 
  -\frac{d_s}{4\sqrt\pi}\frac{ \Gamma (s+1/2)}{\Gamma (s+1)} 
        \zeta_{{\cal B}} (2s , d/2-1) .\nn
\eeq
The same contribution comes from (\ref{naples2f}). The second line in 
(\ref{naples4a}) contributes the $\theta$-dependent piece
\beq
\zeta_\theta (s) &=& \frac{\sin \pi s}\pi 
     \sum d(\nu ) \nu^{-2s} \int\limits_0^\infty dz \,\, z^{-2s}
   \frac \pa {\pa z} \ln \left( 
  1 +\frac{1+t} 2 \left[e^{2\theta } -1 \right] \right) .\label{naples4b}
\eeq
Using standard Taylor series expansion, one first finds
\beq
\frac \pa {\pa z} \ln \left(
  1 +\frac{1+t} 2 \left[e^{2\theta } -1 \right] \right) = 
      -\sum_{l=0}^\infty (-1)^l (\tanh \theta ) ^{l+1} 
\frac z {(1+z^2)^{(l+3)/2} }. \nn
\eeq
This allows the $z$-integrals to be done and an intermediate answer is
\beq
\zeta_\theta (s) = -\frac{d_s} {4\Gamma (s)} \zeta_{{\cal B}} (2s, d/2-1) 
 \sum_{l=0}^\infty (-1)^l (\tanh \theta )^{l+1} \frac
      {\Gamma (s+(1+l)/2)}{\Gamma ((l+3)/2) } .\nn
\eeq
Simplifications occur if $\zeta_{-\theta}$ from (\ref{naples2f}) is 
added,
\beq
\zeta_\theta (s) + \zeta _{-\theta } (s) &=& \frac{d_s} {2\Gamma (s)} 
              \zeta_{{\cal B}} (2s , d/2-1) 
      \sum_{l=0}^\infty (\tanh \theta )^{2l+2} \frac{\Gamma (s+1+l)}
          {\Gamma (2+l)} \nn\\
 &= &\frac {d_s s} 2 \zeta_{{\cal B}} (2s , d/2-1)
 {_2F_1 }(1, s+1, 2 , \tanh ^2 \theta ) \nn\\
  &= & 
       \frac {d_s} 2 \zeta_{{\cal B}} (2s , d/2-1) (\cosh ^{2s} 
               \theta -1) .\nn
\eeq
As expected, for $\theta =0$ the new term $\zeta_\theta (s)+
\zeta_{-\theta} (s)$ vanishes. From
here, the residues $\mbox{Res }\zeta (d/2)$ and $\mbox{Res }\zeta ((d-1)/2)$
are easily determined. The normalization coefficient $a_0$ is confirmed,
for $a_1$ we find
\beq
a_1 = (4\pi ) ^{-(d-1)/2} \frac 1 4 d_s |S^{d-1}|
  \left( \cosh ^{d-1}\theta  -1\right),
\nn
\eeq
with $|S^{d-1}|$ the volume of the $(d-1)$-sphere.
On an arbitrary manifold $\cam$ the coefficient is written in the form
\beq
a_1^\cam = (4\pi )^{-(d-1)/2} \int\limits_{\pa \cam} 
 dy \,\, \delta \,\, \mbox{Tr}( {\bf 1}) ,\nn
\eeq
and we read off
\beq
\delta = \frac 1 4 \left( \cosh^{d-1} \theta -1\right) .\nn
\eeq
As expected, the coefficient $\delta$ depends on $\theta$ as well as 
on the dimension $d$. The dimension dependence is a result of the fact
that the number of $\Gamma$-matrices depends on the dimension $d$. 

\section{Conclusions}
This contribution provides further evidence that special case calculations
play an important role in the determination of the boundary contributions
to the heat equation asymptotics.

Whereas previous applications involved only Bessel functions as special
case solutions, in our first calculation we have provided here an example
where the associated Legendre functions are of relevance. The example involved
was the union of a hemisphere and a ball of radius one, as it turned out to be 
useful in the analysis of transmittal boundary conditions 
\cite{gilk01-601-125}. The choice of a hemisphere simplifies the 
calculations considerably, because the asymptotic behavior 
of the Legendre functions 
needed is relatively simple, see eqs.~(\ref{lege1}) and (\ref{lege2}). If 
instead of the hemisphere a spherical cap is considered, further contributions
will arise because the boundary is not geodesically complete any more and
so $K_{ab} \neq 0$. Presumably, also for this more general situation 
an analysis should be possible,  for example along the lines of 
\cite{barv92-219-201}, where the needed asymptotic behavior of the 
Legendre functions has already been determined. 

The nice feature of using the zeta function for the analysis of the heat
kernel asymptotics is that it also allows for the 
consideration of other spectral functions like the
determinant and the Casimir energy. So as a side effect of the present
calculation an investigation of the influence of edges on these 
quantities can be envisaged.

In the second application presented, we have provided the first
steps into the thorough analysis of the boundary condition 
(\ref{naples2a}). The explicit determination of the heat kernel 
coefficients is in its infancy. A special case calculation is sufficient
to obtain the full $a_1$-coefficient, a result which is already new.
For the higher coefficients, techniques particularly adapted to the 
Dirac operators are needed \cite{gilk00u} and further results 
will be presented 
in \cite{gilk01u}.

{\bf Acknowledgement:} I would like to thank my coauthors M. Bordag, 
T. Branson, G. Cognola, J.S. Dowker, E. Elizalde, G. Esposito, 
P.B. Gilkey, A.Yu. Kamenshchik, J.H. Park and D.V. Vassilevich for 
the very pleasant collaborations on the subject of heat equation 
asymptotics during the course of the 
last years. Special thanks also to the organisers for the invitation 
and for making this wonderful conference possible.


\begin{thebibliography}{10}

\bibitem{abra70b}
M.~Abramowitz and I.A. Stegun,
\newblock {\em Handbook of Mathematical Functions}.
\newblock Dover, New York, 1970.

\bibitem{avra00b}
I.~Avramidi,
\newblock {\em Heat kernel and quantum gravity}.
\newblock Lecture notes in physics m64, Springer-Verlag, Berlin, 2000.

\bibitem{avra91-355-712}
I.G. Avramidi,
\newblock Nucl. Phys., B355 (1991) 712--754.

\bibitem{barn03-19-426}
E.W. Barnes.
\newblock Trans. Camb. Philos. Soc., 19 (1903) 426--439.

\bibitem{barn03-19-374}
E.W. Barnes.
\newblock Trans.~Camb.~Philos.~Soc., 19 (1903) 374--425.

\bibitem{barv92-219-201}
A.O. Barvinsky, A.Yu. Kamenshchik, and I.P. Karmazin,
\newblock Annals Phys., 219 (1992) 201--242.

\bibitem{bord96-37-895}
M.~Bordag, E.~Elizalde, and K.~Kirsten,
\newblock J. Math. Phys., 37 (1996) 895--916.

\bibitem{bord96-182-371}
M.~Bordag, K.~Kirsten, and S.~Dowker,
\newblock Commun. Math. Phys., 182 (1996) 371--394.

\bibitem{bran92-108-47}
T.P. Branson and P.B.~Gilkey,
\newblock J. Funct. Anal., 108 (1992) 47--87.

\bibitem{bran90-15-245}
T.P. Branson and P.B. Gilkey,
\newblock Commun. Part. Diff. Equat., 15 (1990) 245--272.

\bibitem{bran99-563-603}
T.P. Branson, P.B. Gilkey, K.~Kirsten, and D.V. Vassilevich,
\newblock Nucl. Phys., B563 (1999) 603--626.

\bibitem{byts96-266-1}
A.A. Bytsenko, G.~Cognola, L.~Vanzo, and S.~Zerbini,
\newblock Phys. Rept., 266 (1996) 1--126.

\bibitem{camp96-20-1}
R.~Camporesi and A.~Higuchi,
\newblock J. Geom. Phys., 20 (1996) 1--18.

\bibitem{cogn90-241-381}
G.~Cognola, L.~Vanzo, and S.~Zerbini,
\newblock Phys. Lett., B241 (199) 381--386.

\bibitem{dewi65b}
B.S. DeWitt,
\newblock {\em Dynamical Theory of Groups and Fields}.
\newblock Gordon and Breach, New York, 1965.

\bibitem{dewi75-19-295}
B.S. DeWitt,
\newblock Phys. Rep., 19 (1975) 295--357.

\bibitem{dowk94-35-4989}
J.S. Dowker,
\newblock J. Math. Phys., 35 (1994) 4989--4999.

\bibitem{dowk96-13-2911}
J.S. Dowker, J.S. Apps, K.~Kirsten, and M.~Bordag,
\newblock Class. Quant. Grav., 13 (1996) 2911--2920.

\bibitem{dowk99-242-107}
J.S. Dowker, P.B. Gilkey, and K.~Kirsten,
\newblock Contemporary Math., 242 (1999) 107--124.

\bibitem{dowk99-16-1917}
J.S. Dowker and K.~Kirsten,
\newblock Class. Quant. Grav., 16 (1999) 1917--1936.

\bibitem{dowk99-7-641}
J.S. Dowker and K.~Kirsten,
\newblock Communications in Analysis and Geometry, 7 (1999) 641--679.

\bibitem{dowk01-42-434}
J.S. Dowker and K.~Kirsten,
\newblock J. Math. Phys., 42 (2001) 434--452.

\bibitem{eguc80-66-213}
T.~Eguchi, P.B. Gilkey, and A.J. Hanson,
\newblock Phys. Rep., 66 (1980) 213--393.

\bibitem{eliz94b}
E.~Elizalde, S.D. Odintsov, A.~Romeo, A.A. Bytsenko, and S.~Zerbini.
\newblock {\em Zeta Regularization Techniques with Applications}.
\newblock World Scientific, Singapore, 1994.

\bibitem{espo94b}
G.~Esposito.
\newblock {\em Quantum Gravity, Quantum Cosmology and Lorentzian Geometries}.
\newblock Lecture Notes in Physics m12, Springer-Verlag, Berlin, 1994.

\bibitem{full88-310-583}
S.A. Fulling and G.~Kennedy,
\newblock Amer. Math. Soc., 310 (1988) 583--617.

\bibitem{gilk95b}
P.B. Gilkey.
\newblock {\em Invariance Theory, The Heat Equation and the Atiyah-Singer Index
  Theorem, 2nd. Edn.}
\newblock CRC Press, Boca Raton, 1995.

\bibitem{gilk01u}
P.B. Gilkey and K.~Kirsten,
\newblock in preparation.

\bibitem{gilk00u}
P.B. Gilkey and K.~Kirsten,
\newblock Heat asymptotics with spectral boundary conditions II.
\newblock math-ph/0007015.

\bibitem{gilk01naples}
P.B. Gilkey, K.~Kirsten, J.H. Park, and D.V. Vassilevich,
\newblock Asymptotics of the heat equation with 'exotic' boundary conditions or
  with time dependent coefficients.
\newblock International Meeting on Quantum Gravity and Spectral Geometry,
  Naples, Italy, July 2--7, 2001.

\bibitem{gilk01-601-125}
P.B. Gilkey, K.~Kirsten, and D.V. Vassilevich,
\newblock Nucl. Phys., B601 (2001) 125--148.

\bibitem{grad65b}
I.S. Gradshteyn and I.M. Ryzhik.
\newblock {\em Table of Integrals, Series and Products}.
\newblock Academic Press, New York, 1965.

\bibitem{hras84-245-118}
P.~Hrasko and J.~Balog,
\newblock Nucl. Phys., B245 (1984) 118--126.

\bibitem{kirs98-15-5}
K.~Kirsten,
\newblock Class. Quant. Grav., 15 (1998) L5--L12.

\bibitem{avit91-8-1445}
D.M. McAvity and H.~Osborn,
\newblock Class. Quant. Grav., 8 (1991) 1445--1454.

\bibitem{avit91-8-603}
D.M. McAvity and H.~Osborn,
\newblock Class. Quant. Grav., 8 (1991) 603--638.

\bibitem{moss89-229-261}
I.G. Moss and J.S. Dowker,
\newblock Phys. Lett., B229 (1989) 261--263. 

\bibitem{erde55b}
Staff of~the Bateman Manuscript~Project.
\newblock {\em Higher Trascendental Functions}.
\newblock Based on the notes of Harry Bateman, McGraw-Hill Book Company, New
  York, 1955.

\bibitem{olve54-247-328}
F.W. Olver,
\newblock Philos. Trans. Roy. Soc. London, A247 (1954) 328--368.

\bibitem{seel69-91-963}
R.T. Seeley,
\newblock Amer. J. Math., 91 (1969) 963--983.

\bibitem{seel69-91-889}
R.T. Seeley,
\newblock Amer. J. Math., 91 (1969) 889--920.

\bibitem{ven98-15-2311}
A.E.M. van~de Ven,
\newblock Class. Quant. Grav., 15 (1998) 2311--2344.

\bibitem{wipf95-443-201}
A.~Wipf and S.~Durr,
\newblock Nucl. Phys., B443 (1995) 201--232.

\end{thebibliography}
\end{document}